\documentclass[doublespacing]{elsart}

\usepackage{graphicx}
\usepackage{color}
\usepackage{bm}
\def\NOTE#1{{\textcolor{red}{\bf [#1]}}}   
\def\DEL#1{{\textcolor{green}{#1}}}        
\def\ADD#1{{\textcolor{blue}{#1}}}         

\begin{document}
{
%

\def\NOTE#1{{\textcolor{red}{\bf [#1]}}}  
\def\ADD#1{{\textcolor{blue}{#1}}}        
\def\AD#1{{\textcolor{magenta}{#1}}}      
\def\DEL#1{{\textcolor{green}{ #1}}}      
\def\BB#1{{\textcolor{blue}{\bf #1}}}  

\def\hfq{\hfill\quad}
\def\cc#1{\hfq#1\hfq}
\def\dof{d.o.f.}
\def\dofs{d.o.f.}
\def\pt{\PD{}{t}} \def\px{\PD{}{x}}
\def\tvi{\vrule height 12pt depth 5pt width 0pt}
\def\traithorizontal{\noalign{\hrule}}
\def\tv{\tvi\vrule}
\newcommand{\anum}{\lavec{a}}                   
\newcommand{\aop}{\opr{C}}                      
\newcommand{\amat}{\arrfont{C}}                 
\newcommand{\arrfont}[1]{\mbox{\sffamily$\textbf{#1}$}}
\newcommand\assign{\mathbin{:=}}                
\newcommand{\be}{\begin{equation}}
\newcommand{\bfire}{{\it bluefire}}
\newcommand{\bnum}{\lavec{b}}                   
\newcommand{\bvec}[1]{\bf {#1} }                
\newcommand{\conlab}[1]{#1_{\rm{c}}}            
\newcommand{\cpress}{\mathcal{P}}               
\newcommand{\diag}{{\rm diag}} 
\newcommand{\Dmat}{\arrfont{D}}
\newcommand{\DP}[2]{{#1\boldsymbol{\cdot}#2}}   
\newcommand{\DTmat}{\arrfont{D}^{\rm T}}
\newcommand{\ee}{\end{equation}}
\newcommand{\eq}[1]{(Eq. \ref{#1})}
\newcommand{\Fig}[1]{Fig.\ \ref{#1}}
\newcommand{\gaspar}{GASpAR}
\newcommand{\gnum}{\lavec{g}}                   
\newcommand{\grad}{\vec{\nabla}}                
\newcommand{\Hone}{\set{H}^1}                   
\newcommand{\Imat}{\arrfont{I}}
\newcommand{\ipc}[2]{\langle#1,%
#2\rangle}                                      
\newcommand{\ipcsd}[3]{\langle#1,#2\rangle_{\rm #3}}                             
\newcommand{\Jmat}{\boldsymbol\Phi}             
\newcommand{\kraken}{{\it kraken}}
\newcommand{\lavec}[1]{{\boldsymbol#1}}         
\newcommand{\Leg}[1]{{\rm L}_{#1}}              
\newcommand{\Lmat}{\arrfont{L}}                 
\newcommand{\Lop}{\nabla^2}                     
\newcommand{\Ltwo}{{\cal L}_2}                   
\newcommand{\mask}{\boldsymbol{\mathsf{\Pi}}}   
\newcommand{\Mmat}{\arrfont{M}}                 
\newcommand{\opr}[1]{\mathcal{#1}}              
\newcommand{\PD}[2]{\partial_{#2}#1}            
\newcommand{\pdomain}{\set{D}}                  
\newcommand{\pnum}{\lavec{{\mathcal P}}}         
\newcommand{\ppnum}{\lavec{{\mathcal P}}^{+}}    
\newcommand{\pmnum}{\lavec{{\mathcal P}}^{-}}    
\newcommand{\ppmnum}{\lavec{{\mathcal P}}^{\pm}} 
\newcommand{\polynomialsset}[1]{\set{P}_{#1}}   
\newcommand{\pspace}{\set{Y}}                   
\newcommand{\pwisepolysset}[2]{\polynomialsset{#1,#2}}
\newcommand{\Qmat}{\arrfont{A}}                 
\newcommand{\qnum}{\lavec{q}}                   
\newcommand{\Rn}{{\sf {R_v}}}
\newcommand{\Sec}[1]{Section \ref{#1}}
\newcommand{\set}[1]{\mathbf{#1}}               
\newcommand{\setdef}[2]{\left\{#1\left\bracevert#2\right.\right\}}
\newcommand{\si}{\mu}                           
\newcommand{\mtable}[1]{Table \ref{#1}}
\newcommand{\trps}[1]{\raisebox{0pt}{$#1$}^{\rm {\sc{T}}}}
\newcommand{\uh}{\hat{u}}
\newcommand{\union}[2][]{\bigcup_{#2}^{#1}}
\newcommand{\unum}{\lavec{u}}                   
\newcommand{\vnum}{\lavec{v}}                   
\newcommand{\uspace}{\set{U}}                   
\newcommand{\wnum}{\lavec{w}}                   
\newcommand{\uv}[1]{{\vec{e}}^{\:#1}}           
\newcommand{\xnum}{\lavec{x}}                   
\newcommand{\ynum}{\lavec{y}}                   
\newcommand{\znum}{\lavec{Z}}                   
\newcommand{\zpnum}{\lavec{Z}^{+}}
\newcommand{\zpnnum}{\lavec{Z}^{+,n}}
\newcommand{\zmnum}{\lavec{Z}^{-}}
\newcommand{\zmnnum}{\lavec{Z}^{-,n}}
\newcommand{\zpmnum}{\lavec{Z}^{\pm}}
\newcommand{\zmpnum}{\lavec{Z}^{\mp}}

\def\eg{{\it e.g.}\ } 
\def\etal{{\it et al.}} 
\def\ie{{\it i.e.,}\ }
\def\lhs{{\it l.h.s.}\ } 
\def\op{{\it op. cit.}\ } 
\def\resp{{\it resp.}\ }
\def\rhs{{\it r.h.s.}\ } 
\def\rms{{\it r.m.s.}\ } 
\def\viz{{\it viz.}\ }
\def\vs{{\it vs.}\ }
\def\al{Alfv\'en\ }
\def\els{Els\"asser variables\ }
\def\kol{Kolmogorov\ } 
\def\nse{Navier-Stokes equations\ }
\def\u{{\mathbf{u}}} \def\v{\mathbf{v}} \def\x{\mathbf{x}}
\def\dv{\delta {\bf v}}
\def\a{{\vec a}} \def\B{{\vec B}} \def\j{{\vec j}} \def\b{{\vec b}} \def\vu{\vec u} \def\vv{{\vec v}} \def\vx{{\vec x}} \def\Z{{\vec Z}} \def\w{{\vec w}} \def\ztst{{\vec \zeta}}

\newcommand{\curlv} {\nabla \times {\bf v}}
\newcommand{\ba}{\mathbf{a}} \newcommand{\bb}{\mathbf{b}}
\newcommand{\bA}{\mathbf{A}} \newcommand{\bB}{\mathbf{B}}
\newcommand{\Asz}{A_{s_z}}
\newcommand{\alp}{\alpha} \newcommand{\alpm}{\alpha^{-1}} 
\newcommand{\alpmm}{\alpha^{-1}_m} \newcommand{\alpmv}{\alpha^{-1}_v}
\newcommand{\bc}{\mathbf{c}} \newcommand{\bd}{\mathbf{d}}
\newcommand{\bj}{\mathbf{j}} \newcommand{\bk}{\mathbf{k}}
\newcommand{\bom}{\mbox{\boldmath $\omega$}}
\newcommand{\bomp}{\mbox{\boldmath $\omega^+$} }
\newcommand{\bomm}{\mbox{\boldmath $\omega^-$} }
\newcommand{\bompm}{\mbox{\boldmath $\omega^{\pm}$} }
\newcommand{\bu}{\mathbf{u}} 
\newcommand{\bv}{\mathbf{v}}
\newcommand{\bw}{\mathbf{w}}
\newcommand{\bAs}{\mathbf{A_s}} 
\newcommand{\bjs}{\mathbf{j_s}}
\newcommand{\bus}{\mathbf{u_s}} 
\newcommand{\bBs}{\mathbf{B_s}}
\newcommand{\boms}{\mbox{\boldmath $\omega_s$}}
\newcommand{\bx}{\mathbf{x}} \newcommand{\bxp}{\mathbf{x^{\prime}}}
\newcommand{\bzp}{\mathbf{z^{+}}} \newcommand{\bzm}{\mathbf{z^{-}}}
\newcommand{\bzpm}{\mathbf{z^{\pm}}} \newcommand{\bzmp}{\mathbf{z^{\mp}}}
\newcommand{\ca}{{\rm a}} \newcommand{\caa}{{\rm aa}}
\newcommand{\cab}{{\rm a} b} \newcommand{\vba}{vb{\rm a}}
\newcommand{\K}{{\cal K}} \newcommand{\ud}{{\langle{u}^2 \rangle}}
\newcommand{\li}{\ell_{I}} \newcommand{\Rla}{R_{\lambda}}
\newcommand{\up}{{{\bf u}({\bf x})}} \newcommand{\R}{{\cal R}}
\newcommand{\dr}{{\partial_r}} \newcommand{\dt}{{\partial_t}}
\newcommand{\vg}{{{\bf v(x)} \cdot \nabla}}
\setcounter{page}{1}
\setcounter{section}{0}
\begin{frontmatter}
%
%
%
\title{A hybrid MPI-OpenMP scheme for scalable parallel pseudospectral 
computations for fluid turbulence}
\author[ncar:image]{Pablo D. Mininni}
\ead{mininni@ucar.edu}
\author[ncar:image]{Duane Rosenberg}
\ead{duaner@ucar.edu}
\author[psc]{Raghu Reddy}
\ead{rreddy@psc.edu}
\author[ncar:image]{Annick Pouquet}
\ead{pouquet@ucar.edu}
\address[ncar:image]{Institute for Mathematics Applied to Geosciences\\
National Center for Atmospheric Research\\
PO Box 3000, Boulder, Colorado 80307-3000 USA}
\address[psc]{Pittsburgh Supercomputer Center\\
300 S. Craig Street, Pittsburgh, PA 15213 USA}
\begin{abstract}
A hybrid scheme that utilizes MPI for distributed memory parallelism
and OpenMP for shared memory parallelism is presented.  The work is
motivated by the desire to achieve
%
%
exceptionally high Reynolds numbers in pseudospectral
computations of fluid turbulence on emerging petascale, high core-count, massively parallel processing
systems. The hybrid implementation derives from and augments a well-tested scalable MPI-parallelized 
pseudospectral code.
%
%
The hybrid paradigm leads to a new picture for the domain decomposition of the pseudospectral
grids, which is helpful in understanding, among other things, the 3D transpose of the global data that 
is necessary
for the parallel fast Fourier transforms that are the central component of the numerical 
discretizations. Details of the hybrid implementation are provided, and performance tests
illustrate the utility of the method. It is shown that the hybrid scheme achieves near ideal 
scalability up to $\sim 20000$ compute cores with a maximum mean efficiency of 83\%. Data are presented that demonstrate how to choose the optimal
number of MPI processes and OpenMP threads in order to optimize code performance on
two different platforms.

\end{abstract}
\end{frontmatter}
\setlength{\parskip}{1\parskip}	

\section{Introduction}
\label{sec:intro}

Fluid turbulence arises from interactions at all spatial and temporal scales, and is therefore
the quintessential petascale application. The Reynolds number $\Rn$, which measures the strength
of the nonlinearity in turbulent fluid systems, determines the number of degrees of freedom
(\dof) required to resolve all spatial scales, which increases as $\Rn^{9/4}$ (in the Kolmogorov
framework \cite{K41a,K41b}). For geophysical flows,  $\Rn$ is often greater than $ 10^8$, suggesting 
the need to evolve the geo-fluid equations with greater than $10^{18}$ grid points, if completely 
accurate computations of turbulent geophysical flows are to be realized without resorting 
to modeling of unresolved scales. This approach to computing
fluid flows in which all spatial and temporal scales are resolved is called {\it direct numerical 
simulation} (DNS). If the goal is to simulate geophysical flows accurately, such computations
must be carried out at exascale resolutions, which are not currently feasible. But petascale resolutions
are just now becoming available, that can accommodate resolutions of $10^{15}$ grid points, corresponding
to $\Rn \sim 10^7$, which still allows for sufficient scale separation to study 
physically relevant complex turbulent flows.

Pseudospectral methods provide a very useful tool to study the problem because of their computational 
efficiency and high order numerical convergence. Attention is often focused on a $2\pi$--periodic box 
domain in order to study scale interaction
as it allows the use of fast spectral transforms that have a computational complexity of 
$\sim N \log(N)$ instead of $\sim N^2$, 
where $N$ is the linear resolution. For studies of homogeneous and isotropic turbulence, this choice 
is entirely consistent because the domain preserves the underlying translational and rotational 
invariance of the physics. 
But the approach is useful as well for studies of anisotropic or 
inhomogeneous turbulence, which broadens its usefulness.
On the periodic domain, the Fourier basis is optimal, and the pseudospectral
discretization \cite{canuto1988,gottlieb1984,gottlieb1977} is pre-eminent due to the effectiveness of 
the fast Fourier transform (FFT) in converting from configuration to spectral space, and back again.
The pseudospectral method \cite{orszag1972} has thus been used extensively in studies of
computational fluid dynamics (CFD) including turbulence, with references too numerous 
to cite. This method has the extra advantage of accurately capturing the interaction of multiple 
scales with little or no numerical dissipation or dispersion. This is clearly an important property 
for the numerics if we wish to quantify small scale dissipative effects that arise in the context of 
nonlinear turbulent interactions.

Pseudospectral methods, however, require global spectral transforms, and, therefore, are hard to 
implement in distributed memory environments. This has been labeled a crucial limitation of 
the method until domain decomposition techniques arose that allowed computation of serial FFTs 
in different directions in space (local in memory) after performing transpositions. One of these
methods is the 1D (slab) domain decomposition (see e.g., \cite{dmitruk2001}), that enables 
multidimensional FFTs to be parallelized effectively using the Message Passing Interface (MPI). 
However, these methods are often limited in the number of processors that can be used, and 
generalizations to larger processor counts using solely MPI are often expensive or hard to tune 
as transpositions require all-to-all communications. Also, multi-dimensional transforms of 
some non-Fourier basis, such as spherical harmonics, cannot be parallelized using this 
technique. In the present work, a hybrid (MPI-OpenMP) scheme is described that builds upon 
the existing domain decomposition scheme that has been shown to be effective for parallel 
scaling using MPI alone. We leverage this existing domain decomposition method in constructing 
a hybrid MPI-OpenMP model using loop--level OpenMP directives and multi-threaded FFTs. The 
implementation is intended to address several concerns: It addresses the multi---level 
architectures of emerging platforms; and it is also designed to be portable to a variety of 
systems, with the expectation that it will provide scalability and performance without 
detailed knowledge of network topology or cache structure.  

The idea of such loop-level--or implicit--parallelization in concert with MPI is not new. 
To date, these have generally been attempted on small core count systems, and the pure MPI 
scheme is found to outperform the hybrid schemes. In the context of CFD applications, 
it was found that on core counts up to 
256 processors the overall elapsed time (for a finite element solver) was better for the pure MPI 
scheme than for the hybrid, even though the hybrid approach showed improved communication times 
in some cases \cite{yilmaz2007}. A hybrid approach was taken in an implementation of a parallel 3D 
FFT algorithm 
\cite{takahashi2006} that succeeded in reducing the number of cache misses in the algorithm on an 
SMP system. But this approach was again tested only on a small core count platform, and considered 
the FFT algorithm alone, without the full fluid solver. 
To the best of our knowledge, the scheme described herein is the first published implementation of 
a hybrid model in a pseudospectral CFD context that has been attempted on high core count 
systems, and found to scale well.

In the following sections we present a new hybrid implementation. 
We begin first with a description (\Sec{sec:ffttransp})  of the numerical method and the 
underlying domain decomposition scheme. In \Sec{sec:hybrid} the hybrid model is presented, and
a new domain decomposition picture is offered for viewing the distribution of work on
multicore nodes. We also discuss in this section the implementation of the loop--level
parallelization. Benchmarks are provided in \Sec{sec:results}, where we also consider the
overhead and performance of the OpenMP parallelization, and the scalability of the full
hybrid formulation. Finally, in \Sec{sec:conclusion}, we offer some concluding remarks on 
lessons learned and our expectations for future hybrid performance on petascale systems.

\section{The pseudo-spectral method and the underlying domain decomposition}
\label{sec:ffttransp}

All of the work in this paper will be based on simulations of the Navier--Stokes equations:
\begin{equation}
\partial_t {\u} + {\u}\cdot \nabla{\u} = - \nabla p + \nu \nabla^2 {\u} ,
\label{eq_momentum} \end{equation}
\begin{equation}
\nabla \cdot {\u} =0 ,
\label{eq_incompressible}
\end{equation}
where $\u$ is the velocity,  the kinematic viscosity is $\nu$, and the pressure, $p$, can be viewed 
as a Lagrange variable used to satisfy the incompressibility constraint, \eq{eq_incompressible}.
These equations are solved using a pseudo-spectral method \cite{canuto1988,gottlieb1984,gottlieb1977,patterson1971},
in which each component of $\u$ is represented as a truncated (Galerkin) expansion in terms of the Fourier basis, and the
nonlinear term is computed in physical space and then transformed using the fast Fourier transform (FFT), to 
spectral space. The nonlinear term is computed in such a  way that the velocity is projected onto a divergence-free space,
in order to satisfy \eq{eq_incompressible}. Details of this projection and of the dealiasing required by
the action of the nonlinear term are not central to the discussion and can be found elsewhere \cite{canuto1988,patterson1971}, as 
can additional details of the discretization and parallelization of the scheme using solely MPI \cite{gomez2005}.
  
\begin{figure}
\begin{center}
\includegraphics[width=10cm]{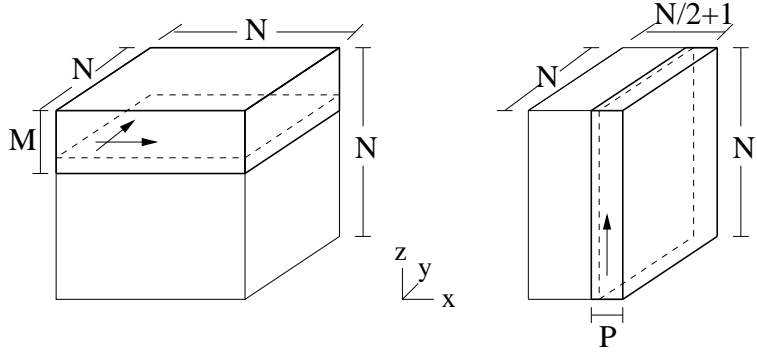}
\end{center}
\caption{Underlying 1D (slab) domain decomposition for pseudo-spectral method ({\it left}). Each processor works on a slab of
size $N\times N\times M$, where $M=N/P$. The FFT is done by first doing the FFTs locally in each slab, in the directions
specified by the arrows, yielding partially transformed data of size  $N/2+1 \times N \times M$. Then, an all-to-all 
communication is done to transpose the data globally ({\it right}), so that the remaining 1D FFT can be done in the direction
specified by the arrow. The data for this step is stored in a cube of size $N/2+1 \times N \times N$, and
each processor now computes the FFT locally in a slab of size $P=(N/2+1)/P$. [Figure adapted from \cite{gomez2005}.]}
\label{fig_slab}
\vspace{.5in}
\end{figure}

The key piece of any pseudospectral method, particularly for parallel computing, is the multidimensional Fourier transform 
algorithm. An efficient parallel implementation of this algorithm is essential for attaining high Reynolds numbers in turbulent 
hydrodynamics simulations, which is of chief concern here. We focus on a 3D Fourier transform of a scalar (or vector component)
field of size $N^3$, with $N$ nodes in each coordinate direction of the $2\pi$-periodic domain. The distribution of real space
points can be viewed as a cubic array of $N^3$ real numbers. In the underlying domain decomposition
each processor receives a ``slab'' of size $N\times N\times M$ node points, where $M = N/N_P$, and $N_P$ is the number
of processors. This is referred to as a 1D domain decomposition because the distribution to processors
occurs in one direction only; this decomposition is visualized in \Fig{fig_slab}. Fourier transforms are performed locally
in the direction of the arrows on the slab owned by a processor. The partially transformed (complex) data resides in a 
cube of size $(N/2+1) \times N \times M$. The reduction in the size of the array results from the fact that a Fourier transform 
of real data $u(x)$ satisfies $\hat{u}(k) = -\hat{u}^*(k)$ (where the asterisk denotes complex conjugate), and therefore only 
half the numbers need to be stored. To compute the (complex) transform in the remaining direction, an all--to--all communication 
is carried out in order to transform the global data cube, and decompose it into slices of size $P\times N\times N$, where 
$P=(N/2+1)/N_P$. Non-blocking MPI communication is used for the all--to--all exchange. This communication allows the transform 
to be carried out in the remaining direction (seen on the right in \Fig{fig_slab}) locally on each processor. Besides 
using non-blocking calls, it is important to make the communication in an ordered way that ensures communication balance. In 
\cite{dmitruk2001}, a list of all possible pairs of MPI tasks is created to this end. Such a list may create problems for 
large processor counts, and as a result here we implement the scheme shown in Fig.~\ref{fig_comm}. Local FFTs are then 
computed using the open source FFTW package \cite{frigo2005,frigo1998}. 

The 1D domain decomposition scheme scales efficiently (\cite{dmitruk2001,gomez2005}), and, when properly implemented, minimizes 
the number of all--to--all communications that must be done to complete the transpose. However, it also limits the number of 
processors to the maximum number of MPI processes that can be used, which is the linear resolution of the run, $N$. In 
practice, departures from linear scaling are often observed before reaching $N$ MPI tasks, as the ratio of computing to 
communication time decreases. We address these issues in \Sec{sec:hybrid}. 

\begin{figure}
\begin{center}
\includegraphics[height=7cm]{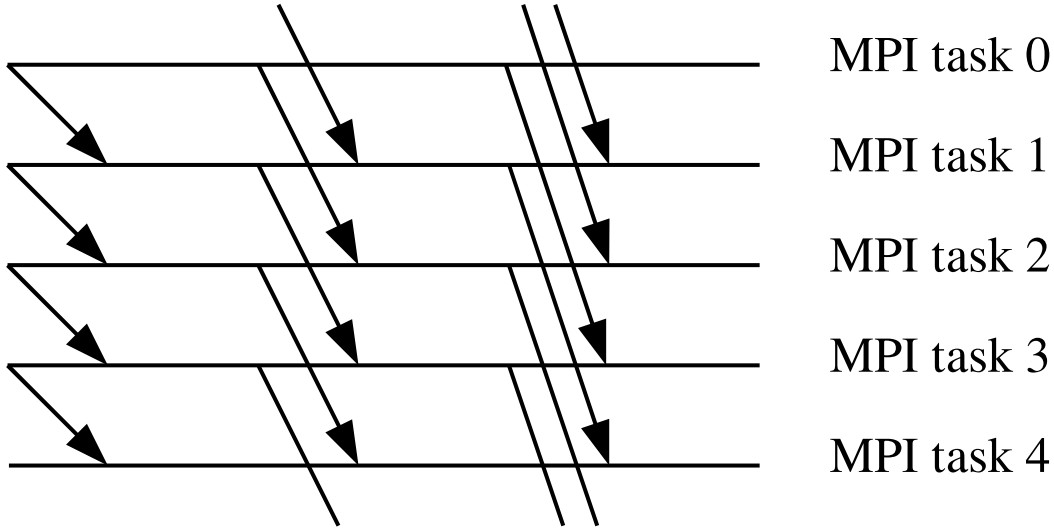}
\end{center}
\caption{Communication pattern for the all-to-all MPI communication to perform the transposition in the parallel FFT. 
Loops are executed in which point-to-point MPI communication (non-blocking send and receive) are performed with increasing 
stride between jobs, until all communications are performed. In the hybrid case, each MPI task can spawn several threads, 
and the communication is handled by the main thread.}
\label{fig_comm}
\vspace{.5in}
\end{figure}

\section{Implementation of the hybrid scheme}
\label{sec:hybrid}

The growing tendency for petascale platforms is toward a hierarchical shared--memory node structure with 
each node having multiple sockets, each with increasing numbers of compute cores with shared or separate
caches, and which may be encapsulated within a non-uniform memory access (NUMA) domain within the node. This 
hierarchical design seems especially suited to a multilevel domain decomposition scheme that can be optimized for 
the hierarchical hardware \cite{hager2009}. In order to address these emerging system designs, and to rectify the
limitation in the underlying slab--only pseudo--spectral domain decomposition strategy of \Sec{sec:ffttransp}, which 
prevents scaling to processor counts beyond the number of MPI processes (linear resolution of the 
problem), we use OpenMP to improve the compute time of each
MPI task. In this scheme, the MPI processes provide a coarse--grain parallelization using the slab domain 
decomposition described above, but OpenMP loop-level constructs and multi-threaded FFTs are applied within each MPI 
job to provide an inner level of parallelization. 

\begin{figure}
\begin{center}
\includegraphics[width=12cm]{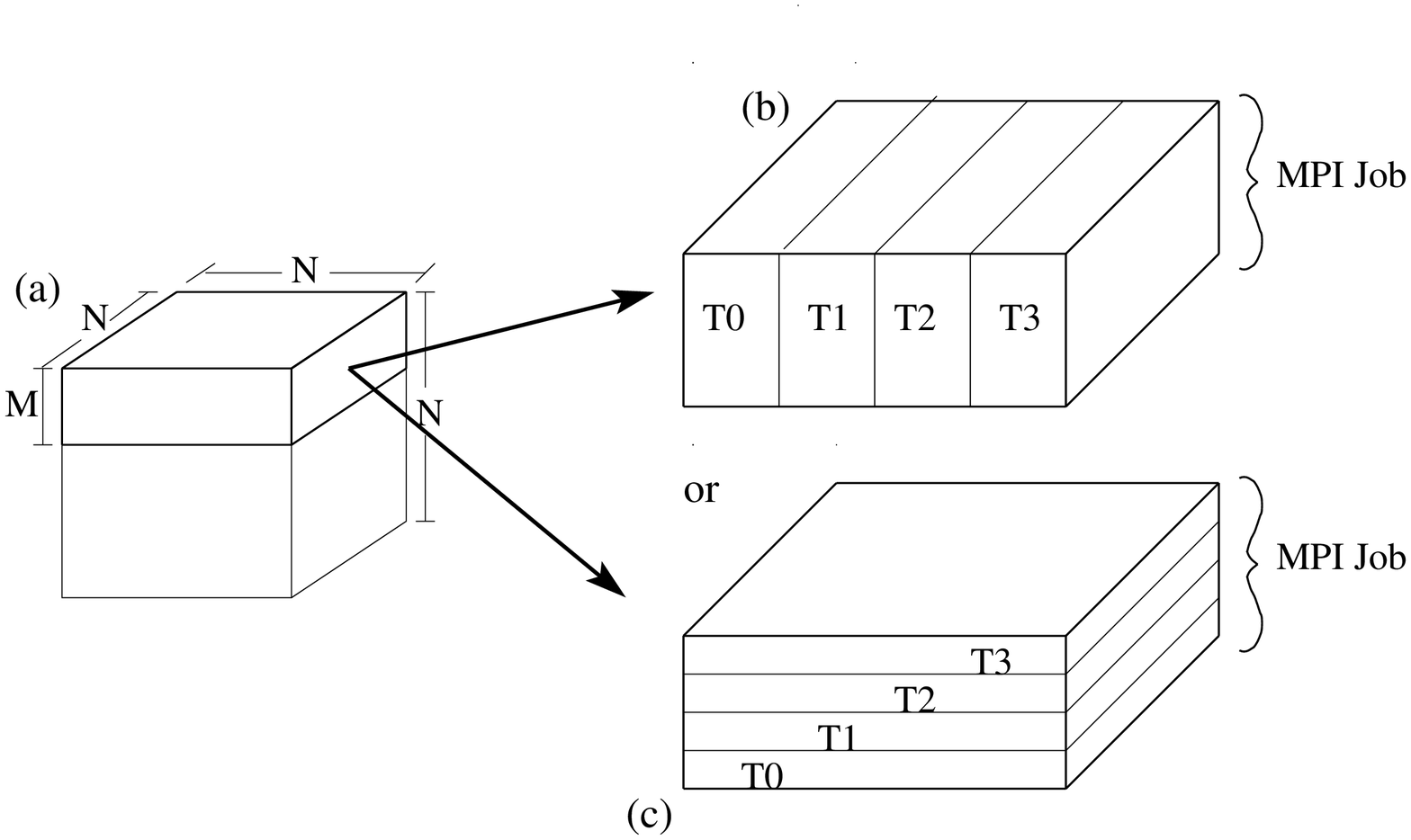}
\end{center}
\caption{Schematic of the new two-level domain decomposition strategy. (a) The 1D domain
decomposition now acts as a coarse--grain MPI-based domain
decomposition step. (b) and (c) A single slab (owned by a single MPI task) is further
parallelized in one of two ways by loop--level OpenMP
directives that distribute different ``chunks'' of the slab to
different threads (here, labeled $T0\ldots T3$) to be worked on, 
speeding up the MPI task. Multi-threaded FFTs are also used in each slab.}
\label{fig_hybrid_dd}
\vspace{.5in}
\end{figure}

Figure \ref{fig_hybrid_dd} illustrates the two--level parallelization scheme. Each MPI task is
parallelized by distributing work among a number of threads ($T0\dots T3$ in the figure), in 
possibly two different ways. This work 
distribution is provided by constructing parallel regions at the loop level using OpenMP directives.
From the point of view of the outer level of parallelization, the multidimensional FFT discussed in 
\Sec{sec:ffttransp} does not change. To show specific inner--level parallelization and to present
the origin of the two different ways to look at the decomposition, we 
provide here a code fragment showing the use of OpenMP directives in carrying out the 
transpose within a slab, crucial for computing the FFT. We focus on this particular algorithm 
because of its importance for the performance of the parallel FFT and also also because it provides a 
good opportunity to highlight an important feature of the code:
\begin{verbatim}
!Multi-threaded FFTs are computed
!All-to-all MPI communication handled by the master thread
!Transposition is now done locally:
!$omp parallel do if ((iend-ibeg)/csize.ge.nthrd) private (jj,kk,i,j,k)
      DO ii = ibeg,iend,csize
!$omp parallel do if ((iend-ibeg)/csize.lt.nthrd) private (kk,i,j,k)
         DO jj = 1,N,csize
            DO kk = 1,N,csize
               DO i = ii,min(iend,ii+csize-1)
               DO j = jj,min(N,jj+csize-1)
               DO k = kk,min(N,kk+csize-1)
                  out(k,j,i) = c1(i,j,k)
               END DO
               END DO
               END DO
            END DO
         END DO
      END DO
\end{verbatim}
Here, the indices $\mathbf{ibeg}$ and $\mathbf{iend}$ indicate the starting and stopping indices
that define the slab for the initial domain decomposition of the data cube. The quantity $\mathbf{csize}$
refers to the cache-size, which is tunable. The outer loop is distributed among threads if the number of 
planes comprising the slab is greater than or equal to the number of threads, $\mathbf{nthrd}$
times the cache size of each thread.  The use of this directive suggests a decomposition
scheme like that illustrated in \Fig{fig_hybrid_dd}(b). If the number of planes is less than 
$\mathbf{nthrd}*\mathbf{csize}$,
then the inner loop is parallelized, which provides a domain decomposition scheme represented 
by \Fig{fig_hybrid_dd}(c).
In this way, we minimize the effect of a potential load imbalance. 

This example not only shows explicitly
how loop--level parallelization is achieved, but also demonstrates one of the ways
in which effective cache utilization is achieved in the local transposition of data by using a technique 
often referred to as ``cache-blocking.'' The three outer loops ensure that the data handled by the 
inner loops is small enough to fit in cache. Since the cache size is tunable, 
this procedure for cache-optimization does not depend on whether the 
thread cache is shared or separate. It has been recognized \cite{hager2009} that the hybrid multi--level
domain decomposition scheme may be especially valuable when taking cache optimization into
account.  All other loops in the code are modified with similar OpenMP 
directives, although most do not need to implement cache-blocking and the $\mathbf{csize}$ dependency. 
As a result, the remaining loops are parallelized as
\begin{verbatim}
!$omp parallel do if ((iend-ibeg).ge.nthrd) private (j,k)
      DO i = ibeg,iend
!$omp parallel do if ((iend-ibeg).lt.nthrd) private (k)
         DO j = 1,N
            DO k = 1,N
!Operations over arrays with indices ordered as A(k,j,i)
            END DO
         END DO
      END DO
\end{verbatim}
The reason for this is that, unlike in the case of the transpose, most of the other loops load 
long lines of contiguous data into cache directly because they have no mixed--index dependencies; 
the transpose requires special treatment because of the dependence of a given block of memory on 
other non-contiguous blocks. Note that in all cases, the loops are ordered--like the above code 
fragment--so that the largest index range keeps the cache lines full. Only a few loops in the 
code (mostly associated with computation of global quantities or spectra which require reductions) 
have to be parallelized using the OpenMP ATOMIC directive.

In both examples, the choice of parallelizing the outer or middle loops based on workload per 
MPI task can be replaced by a COLLAPSE clause in OpenMP 3.0. This clause can be used to parallelize 
nested loops as the ones shown above with only one OpenMP parallel directive. Both solutions have been
benchmarked on different platforms and we observe similar timings. As a result, given the fact 
that the COLLAPSE clause is only available in compilers that support the new OpenMP standard, we 
will use the approach described above in the following examples to ensure portability of the code.

Besides the loop-level parallelism, the FFTs in each slab are also parallelized using the 
multi-threaded version of the FFTW libraries. MPI calls and I/O calls are only executed by the 
main thread in each MPI task. One of the additional benefits of the hybrid scheme presented here 
is that, by reducing the number
of MPI processes, we reduce not only the number of MPI calls, but also 
the amount of data that must be communicated, and hence the size of the MPI buffers 
required to store data. This also allows us to use parallel MPI I/O in environments with 
tens of thousands of cores, as the number of MPI tasks is a fraction of the total number of cores 
used. We will present cases where these considerations become significant in \Sec{sec:results} 
where we provide performance results for the scheme.

\section{Scalability and performance}
\label{sec:results}

A variety of tests have been performed to characterize the overhead, performance and 
scalability of the new hybrid domain--decomposition method. 
Tests were conducted primarily on two platforms: the \bfire\ system at the National Center
for Atmospheric Research (NCAR), and the \kraken\ system at the National Institute for Computational
Sciences (NICS). The \bfire\ platform is an IBM Power 575 system, with $128$ compute nodes, each of which 
contains  $16$ sockets with Power6 processors with $2$ cores each. The compute nodes are interconnected
with InfiniBand; each node has eight 4X InfiniBand double data rate (DDR) links. The \kraken\ system is a 
Cray XT5 with 8256 compute nodes. Each compute node has two six-core AMD Opteron processors 
for a total of 99072 cores. The compute nodes are interconnected with a 3D torus network (SeaStar).
All of the tests discussed here operate in benchmark mode, for which no output other than timings
are produced, and all solve Eqs. \ref{eq_momentum}-\ref{eq_incompressible} for about $50$ timesteps.
Times are measured using the FORTRAN {\bf cpu\_time} routine, and the OpenMP routine {\bf omp\_get\_wtime}.
Timings presented below measure only the average time per timestep for the main time-advance loop; the 
initialization time (including the configuration of FFTW) is not included.

\begin{figure}
\begin{center}
\includegraphics[width=8cm]{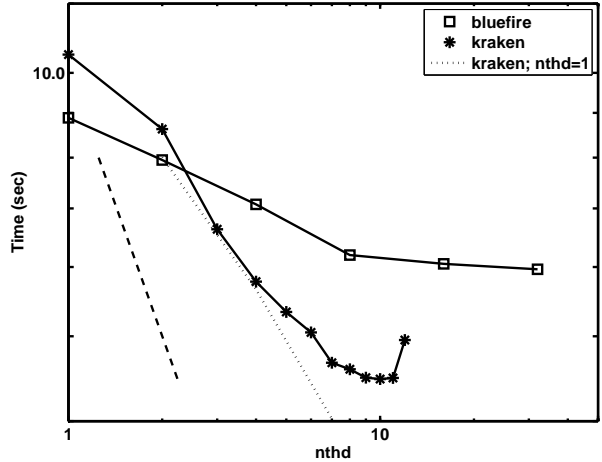}
\end{center}
\caption{Timing results with a single MPI task and multiple threads on two platforms. Both scale roughly the same
for one and 2 threads, but afterward, the \bfire\ jobs run out--of--socket resulting in poor
scaling. The \kraken\ runs scale well up to $7-8$ threads, even though there are only
$6$ cores per socket. The dotted line shows the timings for a single thread, while varying the
number of MPI processes (hence, the $\mathbf{nthd}$ axis refers to the number
of MPI tasks for this curve only). 
The dashed line represents ideal scaling, and is also used in all subsequent scaling
plots.}
\label{fig_1mpi_manyth_kr256}
\vspace{.5in}
\end{figure}


In the first series of tests, we consider the overhead and performance of OpenMP. 
The first test thus considers a single MPI process, and variable number of threads 
{\bf nthd} with a fixed linear resolution of $N=256$. The results are presented in 
\Fig{fig_1mpi_manyth_kr256}. The performance for $1$ and $2$ threads is comparable for both platforms. 
After this, \bfire\ communicates out--of--socket, and its scaling decreases. We expect
that as the core counts increase for this platform (\eg, as for the Power7 system) this 
problem will not be as severe. For \kraken, there are 
$6$ cores per socket, but we still see very good scaling to about $7$ threads. Moreover, 
the departures from the ideal scaling observed in \kraken\ while computing in-socket 
seem to be associated with the hardware (e.g., with saturation of the memory bandwidth) and 
not specifically with OpenMP. This we conclude from tests in which OpenMP parallelization is turned
off, and the number of MPI tasks is increased.
In using from 1 to 6 MPI tasks, the same scaling is observed as with pure OpenMP 
parallelization (\Fig{fig_1mpi_manyth_kr256}). 

\begin{figure}
\begin{center}
\includegraphics[width=6.5cm]{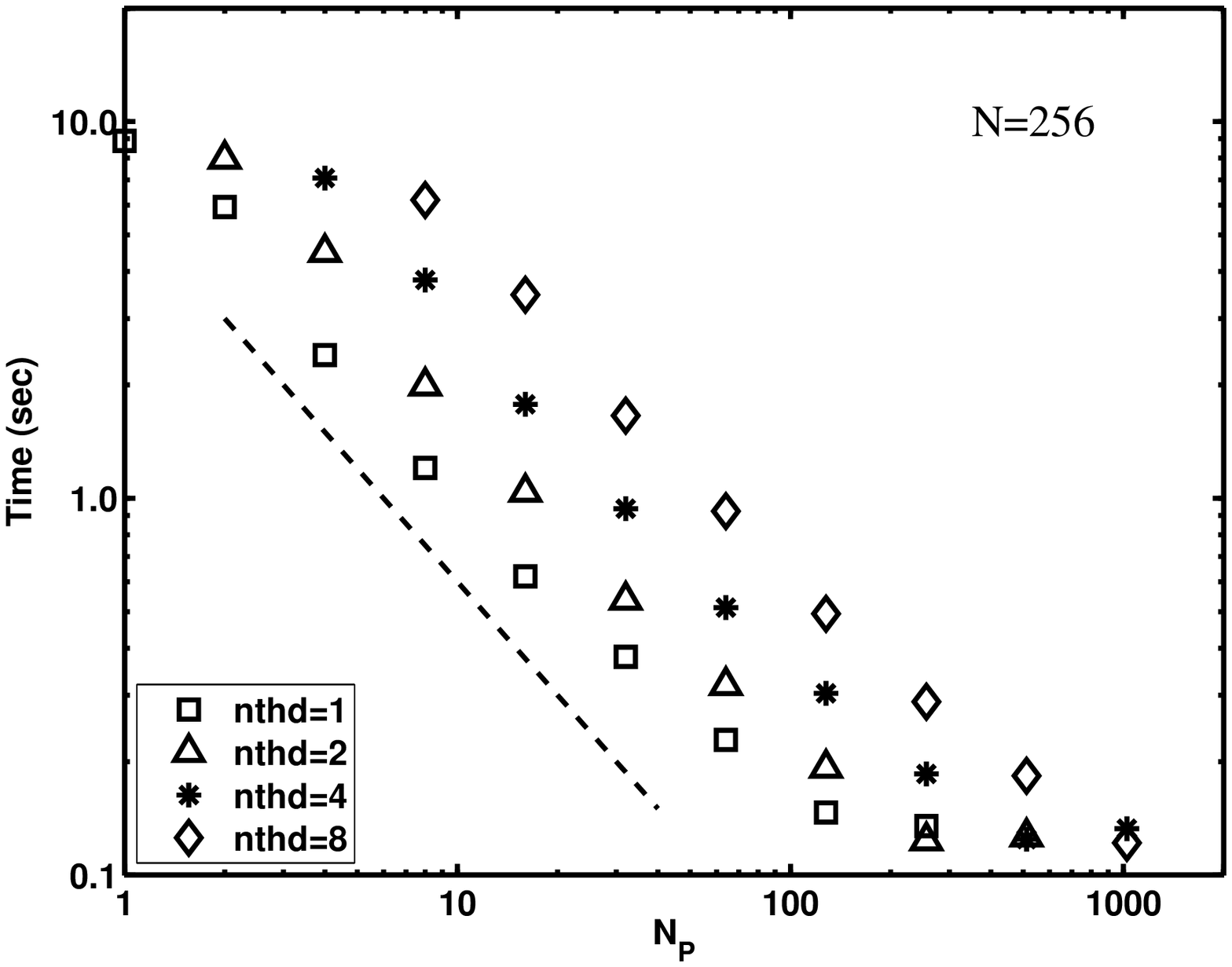}
\includegraphics[width=6.5cm]{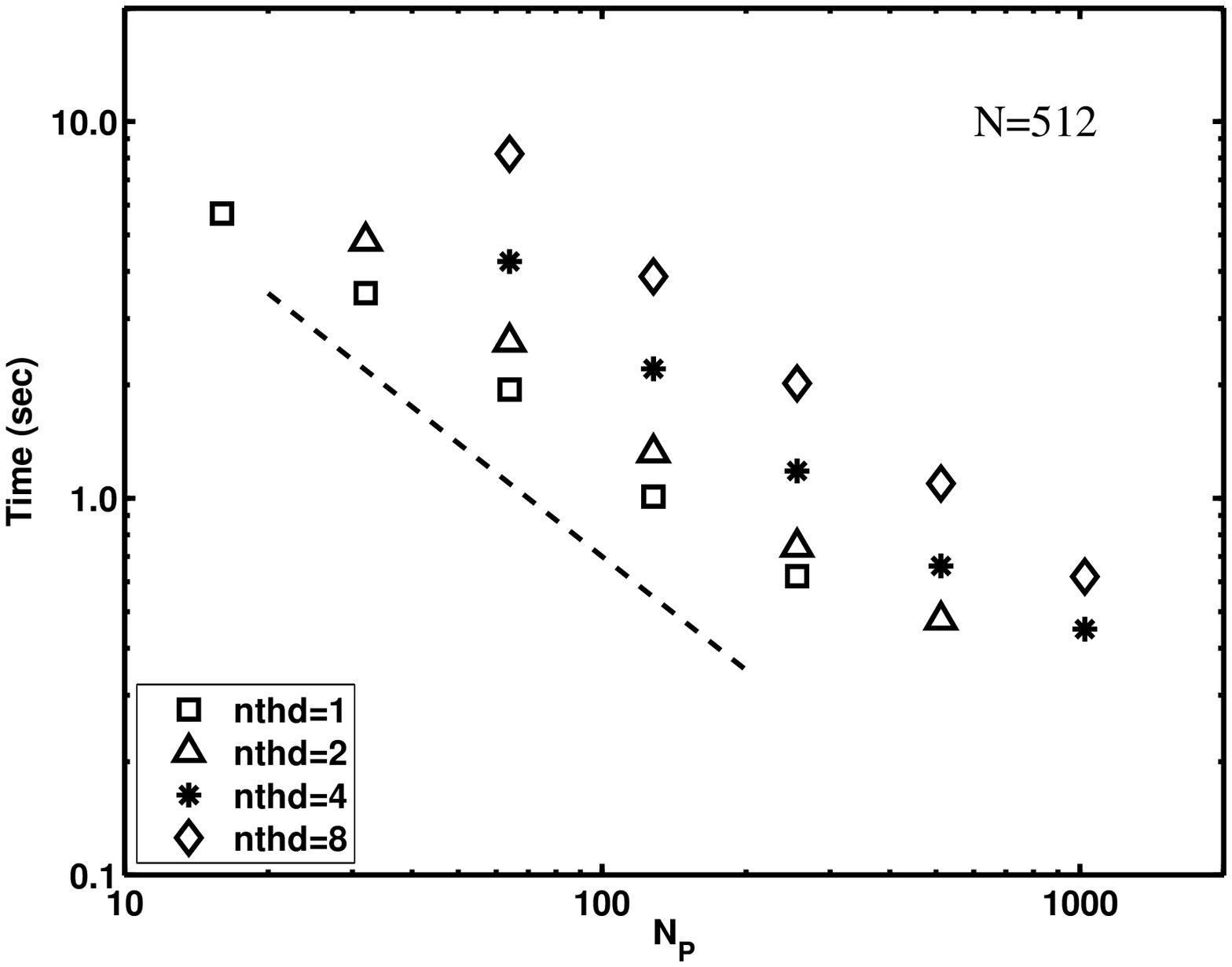}
\end{center}
\caption{Two sets of \bfire\ runs, for $N=256$ ({\it left}) and for $N=512$ ({\it right}). The curves
represented by the different symbols are runs at a constant number of threads, as given in the
legend. Note, in particular, that the difference in run time between the $1$- and $2$-thread
surveys are smaller for the runs with $N=512$, than for the cases where $N=256$,
which suggests that the thread overhead will manifest itself with smaller work load on this 
platform. Also note tje almost linear scalability up to 1000 processors.}
\label{fig_manyth_bf}
\vspace{.5in}
\end{figure}

In order to examine effects of OpenMP overhead on \bfire\ results more closely, we compare two runs
at different resolutions, one at $N=256$, and one at $N=512$ for a series of thread counts. 
These results are given in \Fig{fig_manyth_bf}. In each plot the symbols refer to the same 
$\mathbf{nthd}$, and $N_P$ is varied by changing the number of MPI processes. The MPI tasks were 
bound to processors (using ``processor binding''), and symmetric multi-threading (SMT) was
disabled. The first observation 
is that, for $N=256$, the gains as $\mathbf{nthd}$ is increased (for any fixed number of MPI tasks) 
are roughly the same as the ones reported in \Fig{fig_1mpi_manyth_kr256} for only 1 MPI task. 
However, for large numbers of MPI tasks, using $\mathbf{nthd}=2$ gives better timings than 
$\mathbf{nthd}=1$ using the same total number of processors (e.g., compare the triangle and the 
square at $N_P=256$ with the square at $N_P=128$). Increasing the number of threads further 
does not give substantial speed-ups. This is observed more clearly in the $N=512$ runs. 
In this case, the slope between runs with $\mathbf{nthd}=1$ and $\mathbf{nthd}=2$ is larger, 
indicating better gains as the size of the problem is increased.

\begin{table}
\caption{Efficiency of runs with $N$ linear resolution in \bfire, taking as reference runs with 
$N_{P_0}=N/2$ cores and $\mathbf{nthd}=1$ threads.}
\label{MPIvsOpenMP}
\vspace{.3in}
\begin{center}
\begin{tabular}{lcc}
\hline
$N$  & $N_P=N$, $\mathbf{nthd}=1$ & $N_P=N$, $\mathbf{nthd}=2$ \\
\hline
256  &     $0.54$                 &     $0.59$                 \\
512  &     $0.58$                 &     $0.65$                 \\
1024 &     $0.63$                 &     $0.66$                 \\
\hline
\end{tabular}
\end{center}
\vspace{.5in}
\end{table}

As a result, in \bfire\ there appears to be an effect due to the thread overhead that is noticeable 
when using 2 threads (in-socket) and the problem size is small: we see that the differences in run 
time between the $1$-thread and $2$-thread runs is smaller as resolution is increased. Then, as the 
threads are out-of-socket, extra overhead appears (although for fixed number of threads, very good 
scaling is found with increasing number of MPI tasks). This can be further observed considering runs 
with $N=1024$. \mtable{MPIvsOpenMP} shows the efficiency
\begin{equation}
\epsilon = \frac{N_P  T}{N_{P_0} T_0} ,
\end{equation}
where $T$ is the time per time step, and $N_{P_0}$ and $T_0$ are respectively the number of cores 
and times measured in a reference run; we consider $N_{P_0}=N/2$ with $\mathbf{nthd}=1$ as the 
reference run. For a fixed number of processors $N_P=N$, the efficiency is best if two threads 
are used instead of one, and as resolution is increased efficiency improves. If $N_P=2N$ and four 
threads are used, efficiency also increases but is at most $\approx 0.4$ for $N=1024$.

\emph{These results suggest that a hybrid approach may be most useful for large enough simulations in 
environments with large processor counts and when a large number of cores is available in the same 
socket.} To verify this we consider the scaling to high core counts on \kraken. For these 
runs, we set $N=1536$, $N=3072$, and $N=4096$, with $\mathbf{nthd}=6 \;{\rm or }\; 12$. At these 
resolutions, simulations with 1 thread cannot be executed as there is not enough memory per core 
in \kraken\ to allocate the arrays. Several simulations at lower resolution ($N=512$) were done to 
explore configuration parameters. This mainly involved NUMA options in the compiler (PGI), different 
binding configurations, MPI environment settings, and distribution of jobs among processors. We 
observed no substantial differences in the timings when changing the job distribution. 
Binding processors when $\mathbf{nthd}=6$ using the run command instead of NUMA options in the 
compiler was found to be best, although by a small margin ($\approx 5\%$). The implementation 
of MPI on \kraken\ can also be configured to do fast copies in memory of the data when sending 
and receiving large messages. This gives a substantial speed up of the code (8 -- $10\%$) 
but was found to require large amounts of memory that created problems with the largest 
resolutions. As a result, to compare on an equal footing, the 
runs described below were compiled with -O2 optimization, without using fast memory copy in MPI, 
and using the run command
\begin{verbatim}
aprun -n $NMPI -S 1 -d $OMP_NUM_THREADS executable
\end{verbatim}
for the $\mathbf{nthd}=6$ runs, and the run command
\begin{verbatim}
aprun -n $NMPI -d $OMP_NUM_THREADS executable
\end{verbatim}
for $\mathbf{nthd}=12$. In the former case, the {\rm -S 1} option tells {\rm aprun} to bind one 
MPI process per socket, and {\rm NMPI} refers to the total number of MPI processes. It should be 
noted that in the runs for which the most aggressive optimization options can be used (e.g., if 
enough memory is available), improvements in the times of up to $20\%$ were found.

\begin{figure}
\begin{center}
\includegraphics[width=8cm]{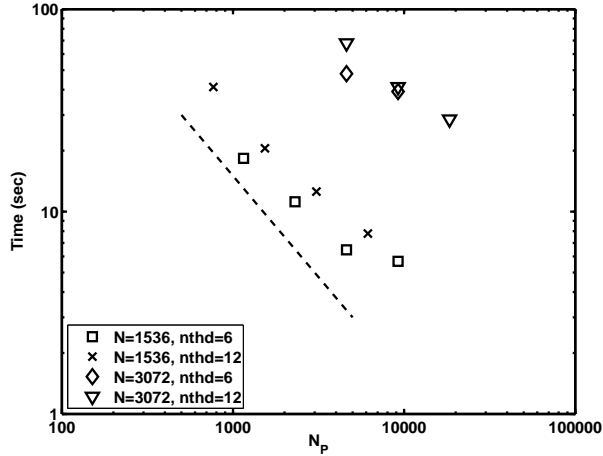}
\end{center}
\caption{Scalability timings for two sets of runs on \kraken. The squares and crosses represent the timings
for a $N=1536$ run using $6$ and $12$ threads, respectively. The triangles, and diamonds represent the same
for a run size of $N=3072$. Note the cross--over in performance at about $N_P=10000$ where the $\mathbf{nthd}=12$
configuration outperforms the $\mathbf{nthd}=6$ configuration for both problem sizes.}
\label{speed_hyb_kraken}
\vspace{.5in}
\end{figure}

The results are given in \Fig{speed_hyb_kraken}. We see that good speedup is achieved up
to $\sim 20000$ cores, but there are some interesting observations. The mean parallel efficiencies 
for these runs are $76$, $83$, $61$, and $71\%$, respectively, from top to bottom in the legend of 
\Fig{speed_hyb_kraken}. Maximum efficiencies observed are slightly above 1, 
indicating super-scaling for some configurations. The efficiency measured for the jobs with larger processor 
count is given in \mtable{kraken_eff}. Based on the results shown in \Fig{fig_1mpi_manyth_kr256}, we 
can expect optimal results for $\mathbf{nthd}=6$. However, the runs with $\mathbf{nthd}=12$ scale 
uniformly better than the $\mathbf{nthd}=6$ runs. Moreover, from the point of view of actual timings the 
$\mathbf{nthd}=6$ cases perform better up to a certain point at which the $\mathbf{nthd}=12$ runs outperform them; 
this point appears to occur at about the same $N_P$ for both sets of runs. This is particularly clear for the 
run with $N=3072$, where the same times are measured for $N_P \sim 10000$ using $\mathbf{nthd}=6$ and 
$\mathbf{nthd}=12$. The results are consistent with the findings in \bfire, but the larger processor count and 
cores-per-socket in \kraken\ allow us to obtain significant gains using the hybrid approach in the latter case. 
We conclude that if the workload per MPI process becomes too small, it is better to use more threads even if 
this puts threads out-of-socket.

\begin{table}
\caption{Efficiency of runs with $N=3072$ linear resolution in \kraken, taking as reference runs with 
$N_{P_0}=N_P/2$ cores and same number of threads.}
\label{kraken_eff}
\vspace{.3in}
\begin{center}
\begin{tabular}{ccc}
\hline
$N_P=9216$, $\mathbf{nthd}=6$ & $N_P=9216$, $\mathbf{nthd}=12$ & $N_P=18432$, $\mathbf{nthd}=12$ \\
\hline
       $0.58$                 &     $0.61$                    &     $0.72$                      \\
\hline
\end{tabular}
\end{center}
\vspace{.5in}
\end{table}

\section{Discussion and conclusion}
\label{sec:conclusion}

We have presented a hybrid MPI-OpenMP model for a pseudo--spectral CFD code. Beginning with
an underlying ``slab'' domain decomposition adequate for parallelization by MPI, we have 
shown how the basic method is modified by
loop--level parallelization to create a two--level parallelization scheme. The new level
of parallelization can be thought of as modifying the underlying domain decomposition
scheme, and we have pointed out precisely how this has been done depending on the size of the 
problem, number of threads, and number of MPI tasks.

The hybrid code has been tested primarily on two systems: the IBM Power6 system \bfire\ at NCAR, and the
Cray XT5 system \kraken\ at NICS.  We have tested the thread overhead and performance, and found
limitations of small socket core counts in \bfire. We have also discovered that there is a resolution
threshold, $N$, below which the thread overhead manifests itself more clearly on \bfire\ and reduces 
scalability. In terms of large core counts, our results show good scalability up to about $20000$ 
processors on the \kraken\ system. For large enough problems, we find the best 
scalability when the number of threads is $12$ (one MPI process per compute node). On the other
hand, we find that the performance time is better 
when $\mathbf{nthd}=6$, until the workload per MPI process is large enough, 
at which point, the performance time is better for the case where $\mathbf{nthd}=12$.
We find that, for a given MPI/OpenMP configuration, and a given resolution, the results are
consistent from run--to--run, with little fluctuation in terms of scalability or run time. 

Our experimentation has suggested a number of ways in which to improve the compute time of
\kraken\ runs. Perhaps the most important of these involve configuring the MPI environment.
For the large message sizes we are using, setting the MPI environment
variable {\rm  MPICH\_FAST\_MEMCPY} yields an $8-10\%$ speedup over runs that
do not use it, but it requires significantly more buffer memory. This increase in buffer requirements
can prevent the code at large resolutions from fitting into memory, and must be considered
carefully before attempting a production run. As an example, for $N=4096$ using this configuration, 
we could only execute the code using 24576 processors and 6 threads, and any other distribution using 
the same or smaller number of processors and changing the number of threads failed because of 
insufficient memory. We have not attempted larger resolution runs yet, but we note that the memory 
issue addressed here will become more of a concern the larger $N$ becomes.





The hybrid scheme introduced here is not the only way in which to decompose the pseudo--spectral
grid. An alternative is to retain a pure MPI model as in \cite{yeung2005}. In this model, the domain
decomposition takes the form of ``pencils'' which yield  a 2D ($N^2$) distribution among MPI processes, 
and OpenMP is not required. This technique is also found to scale well \cite{donzis2008} to large core counts,
although severe fluctuations in performance are observed even within a given processor--domain mapping.
The pure MPI model does not suffer from effects of thread overhead (thread re-starts and
synchronization) that we observe in smaller resolution runs, nor from potential problems with
compiler optimizations that may arise when OpenMP is used \cite{hager2009}.
Nevertheless, the hybrid method described here can be applied to non--Fourier basis spectral methods which may
be impossible to parallelize with the 2D distribution (\eg, spherical harmonics). 
As pointed out in \Sec{sec:hybrid}, our hybrid method offers a two--level parallelization method that may be
more effective in mapping the domain to the hierarchical architectures that are now emerging, and better 
suited for environments with multiple cores per socket. Indeed, as noted in \Sec{sec:results}, 
based on our tests on \bfire\ and \kraken\ we expect the hybrid approach to provide better
performance results in coming years, as the number of cores per socket continues to increase.
The hybrid scheme may also aid in the MPI 
memory problems mentioned above, in that fewer MPI processes require less buffer memory. 
We intend in the future to continue testing this method to higher resolution as accessibility to a larger number of processors becomes more readily available. Since the code described here integrates the Navier-Stokes or magnetohydrodynamics equations when coupling to a magnetic field, including rotation, the hybrid scheme we have developed will prove useful in a variety of geophysical and astrophysical phenomena. 
And finally, we note that this approach works well even if the aspect ratio of the computational domain is not equal to unity.


\begin{ack}
Computer time was provided by NSF under sponsorship of the National
Center for Atmospheric Research, and under TeraGrid (project number ASC090050) 
and is gratefully acknowledged. 
\end{ack}

\bibliographystyle{plain}
\bibliography{dlr}

\end{document}